\documentclass[12pt,preprint]{aastex} 
\usepackage{psfig,natbib,amsmath}

\newcommand{\citepeg}[1]{\citep[{e.g.,}][]{#1}}

\def\lsim{\hbox{ \rlap{\raise 0.425ex\hbox{$<$}}\lower 0.65ex\hbox{$\sim$}}}
\def\gsim{\hbox{ \rlap{\raise 0.425ex\hbox{$>$}}\lower 0.65ex\hbox{$\sim$}}}
\def\arcmin{\hbox{$^\prime$}}
\def\arcsec{\hbox{$^{\prime\prime}$}}
\def\fd{\hbox{$~\!\!^{\rm d}$}}
\def\fh{\hbox{$~\!\!^{\rm h}$}}
\def\fm{\hbox{$~\!\!^{\rm m}$}}
\def\fs{\hbox{$~\!\!^{\rm s}$}}
\def\ale{\mathrel{\hbox{\rlap{\hbox{\lower4pt\hbox{$\sim$}}}\hbox{$<$}}}}
\def\age{\mathrel{\hbox{\rlap{\hbox{\lower4pt\hbox{$\sim$}}}\hbox{$>$}}}}

\begin{document}

\title{A Putative Early-Type Host Galaxy for GRB 060502B: Implications for the Progenitors of Short-Duration Hard-Spectrum Bursts}

\def\berk{1}
\def\chic{2}
\def\lick{3}
\def\harvard{4}

\author{
        J. S. Bloom\altaffilmark{\berk}, 
        D. A. Perley\altaffilmark{\berk},
        H.-W. Chen\altaffilmark{\chic}, 
        N. Butler\altaffilmark{\berk}, \\
        J. X. Prochaska\altaffilmark{\lick},
        D. Kocevski\altaffilmark{\berk},
        C. H. Blake\altaffilmark{\harvard},
        A. Szentgyorgyi\altaffilmark{\harvard},
        E. E. Falco\altaffilmark{\harvard},
        D. L. Starr\altaffilmark{\berk}
}

\affil{$^\berk$ Department of Astronomy, 601 Campbell Hall, 
        University of California, Berkeley, CA 94720-3411.}

\affil{$^\chic$ Department of Astronomy \& Astrophysics, University of Chicago, Chicago, IL 60637.}

\affil{$^\lick$ University of California Observatories/Lick
Observatory, University of California, Santa Cruz, CA 95064.}

\affil{$^\harvard$ Harvard-Smithsonian Center for Astrophysics, 60 Garden Street, Cambridge, MA 02138.}

\begin{abstract}

Starting with the first detection of an afterglow from a
short-duration hard-spectrum $\gamma$-ray burst (SHB) by Swift last
year, a growing body of evidence has suggested that SHBs are
associated with an older and lower-redshift galactic population than
long-soft GRBs and, in a few cases, with large ($\age 10$ kpc)
projected offsets from the centers of their putative host galaxies.
Here we present observations of the field of GRB\, 060502B, a SHB
detected by Swift and localized by the X-ray Telescope (XRT). We find
a massive red galaxy at a redshift of $z=0.287$ at an angular distance
of 17.1\arcsec\ from our revised XRT position. Using associative and
probabilistic arguments we suggest that this galaxy hosted the
progenitor of GRB\, 060502B. If true, this offset would correspond to
a physical displacement of $73 \pm 19$ kpc in projection, about twice
the largest offset inferred for any SHB to date and almost an order of
magnitude larger than a typical long-soft burst offset.  Spectra and
modeling of the star-formation history of this possible host show it
to have undergone a large ancient starburst.  If the progenitor of
GRB\, 060502B was formed in this starburst episode, the time of the GRB
explosion since birth is  $\tau \approx 1.3 \pm 0.2$ Gyr and the minimum 
kick velocity of the SHB progenitor is $v_{\rm kick, min} = 55 \pm 15$ km s$^{-1}$.

\end{abstract}


\section{Introduction}

Since the seminal work of \citet{kmf+93}, a consensus view has emerged
that short-duration hard-spectrum GRBs (SHBs) arise from a separate
physical population than long-duration soft-spectrum GRBs (LSBs). The
populations are distinguished phenomenologically by an observed
bimodality in the GRB duration distribution \citep{mgi+81,ncdt84} and
an apparent corresponding bimodality in spectral hardness. While most
LSB progenitors are now believed to be due to the death of massive
stars, without a successful detection of an afterglow or a host galaxy
the nature of the SHBs remained a mystery until recently.

In May 2005, the Swift satellite detected and localized SHB\, 050509B
and, for the first time, found a fading X-ray afterglow
\citep{gso+05}; this was the first SHB localized quickly ($\ale 10$ s)
and accurately ($< 100$ arcsec$^{2}$). Ground-based followup
observations led to the discovery of an early-type galaxy at a
redshift of $z=0.258$ approximately 10\arcsec\ from the X-ray
afterglow position \citep{bpp+06}.  A chance association with such a
galaxy was deemed unlikely even under conservative assumptions ($P <$
few percent) and stood in stark contrast with the lines-of-sight of
LSBs, with which no association of with an early-type was ever
made. Both the nature of the burst itself (lacking any supernova
signature; \citealt{hsg+05}) and the location (in the halo of a red
galaxy with very little star formation) suggested a progenitor of a
very different nature from the purported progenitors of LSBs.  In
particular, these observations were in close agreement with
predictions \citep{bp95,bsp99,fwh99} for the nature of the environment
-- particularly, the offset from host galaxy and the type of the host
-- associated with the merger of a degenerate binary
(e.g.~\citealt{npp92}).

Further Swift and {\it HETE-2} detections of SHBs have continued to support this
hypothesis, though SHBs are not universally at large offsets and are not always
associated with early-type galaxies (see \citealt{bp06} for a review).
SHB\, 050724 \citep{bpc+05,pbc+06,gcg+06} and 050813 \citep{pbc+06},
like 050509B, were found to be in close association with old, red
galaxies (see also \citealt{ltf+06}).  SHB\, 050724 had optical and radio afterglow emission that
pinpointed its location to be within its red host, making the
association completely unambiguous, though the association of 050813
with any single host remains somewhat tentative.  Not all hosts lack
active star formation; SHB\, 050709
\citep{vlr+05,hwf+05,ffp+05,cmi+06} and 051221A \citep{sbk+06} both
had optical afterglows and were associated with galaxies with evidence
for current star formation.  However, despite the availability of both
X-ray and optical afterglow locations, no nearby host has successfully
been identified for either SHB\, 060121 or SHB\, 060313 (although see
\citealt{hjo+06}).

In this article we examine the field of Swift SHB\,060502B
\citep{tbb+06} and, in \S \ref{sec:0502b}, we present imaging and
spectroscopy of a bright red galaxy near the X-ray afterglow
position. In \S \ref{sec:host} we present evidence that supports the
notion that the progenitor of SHB\,060502B was born in that
galaxy. Accepting this connection we discuss the implications of the
nature of the host and offset for the progenitors of SHBs. Though the
association of this galaxy with the GRB is the most tenuous of
SHB--host associations thus far proposed, we conclude in \S
\ref{sec:conc} that there are both observational and theoretical
motivations to accept this association for this and (similarly
configured) future SHBs. Some of our work on this GRB was given
preliminarily in \citet{bpk+06}; our results presented herein are
consistent with, but supersede that reference.  Throughout this {\it
Letter} we assume $H_0 = 71~h_{71}$ km s$^{-1}$ Mpc$^{-1}$,
$\Omega_m = 0.3$, $\Omega_\Lambda = 0.7$.

\section{SHB 060502b and $G^*$}
\label{sec:0502b}

At 2006 May 02 17:24:41 {\sc UTC}, the Swift Burst Alert Telescope
triggered \citep{tbb+06} on a GRB consisting of a strong single-spike
with a FWHM of $40$ ms and a possible second precursor spike; ninety
percent of the total fluence arrived over a timespan of $90 \pm 20$
milliseconds \citep{sbb+06}, making it one of shortest GRBs localized
by Swift (E.~Troja, private communication). The X-ray afterglow was
localized to a final position of $\alpha$ = 18\fh35\fm45\fs.74,
$\delta$ = +52\fd 37\arcmin\ 52\arcsec.47 (J2000) with a 4.4\arcsec\
uncertainty radius (90\% confidence) \citep{tbg06}. Using 7 X-ray
persistent sources found within 10\arcsec\ of 8 sources in the Digitized Sky Survey
near the X-ray positions, we find a consistent position of $\alpha$ =
18\fh35\fm45\fs.48, $\delta$ = +52\fd 37\arcmin\ 52\arcsec.7 (J2000)
with a 4.36\arcsec\ uncertainty radius (90\% confidence); this accounts for the small shift of the DSS astrometric frame to the (more precise) 2MASS frame\footnotemark\footnotetext{An outline of the XRT reanalysis technique was presented in \citet{bb06}. Details may be found at {\tt http://lyra.berkeley.edu/$\sim$nat/Swift/xrt\_astrom.html}}.  Starting 74
seconds after the GRB, the Ultraviolet-Optical Telescope (UVOT)
onboard Swift obtained a deep unfiltered exposure of 100 sec and found
no optical afterglow candidate to a limiting magnitude of 19.1\, mag
\citep{tbb+06}. Likewise, no optically variable counterpart was found
in rapid groundbased imaging to $R < 20$\, mag several minutes to
hours after the GRB \citep{lkt+06,zqw+06,kkf06,tuy+06,mvo+06,hm06}. No
variable optical counterpart was found in deep image differencing of
r$^\prime$ GMOS/Gemini 8-m data taken at 0.7 and 1.7 days after the GRB
\citep{pbf+06}. Three sources in the refined Swift XRT error circle
were identified, one of which was shown through spectroscopy to be a
Galactic star \citep{bcr06,hm06a,rks+06}.  Three additional sources
are located in or near our modified XRT error circle.

\subsection{Imaging}

\begin{figure*}[p] 
\centerline{\psfig{file=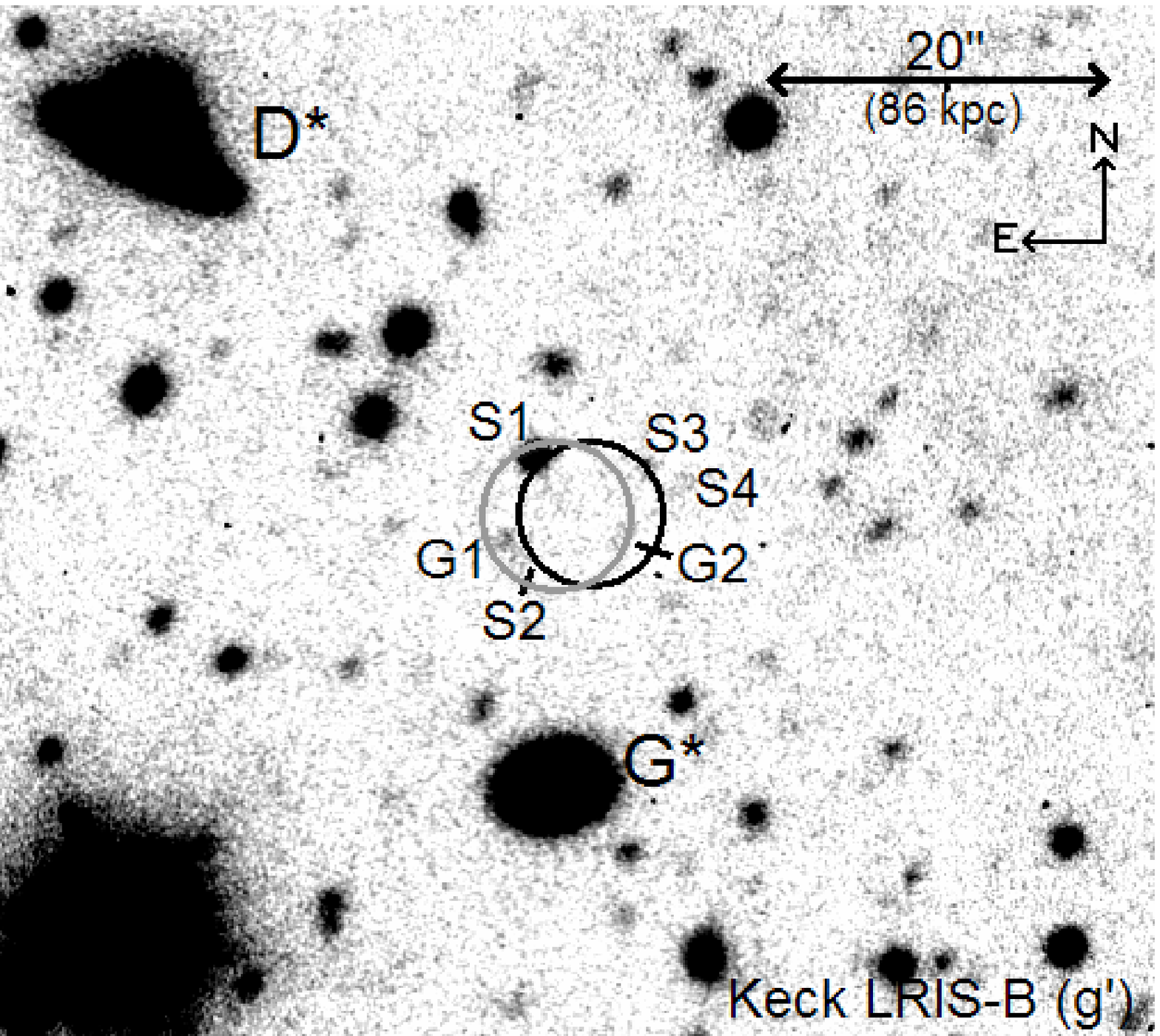,width=4.0in,angle=0}}
\centerline{\psfig{file=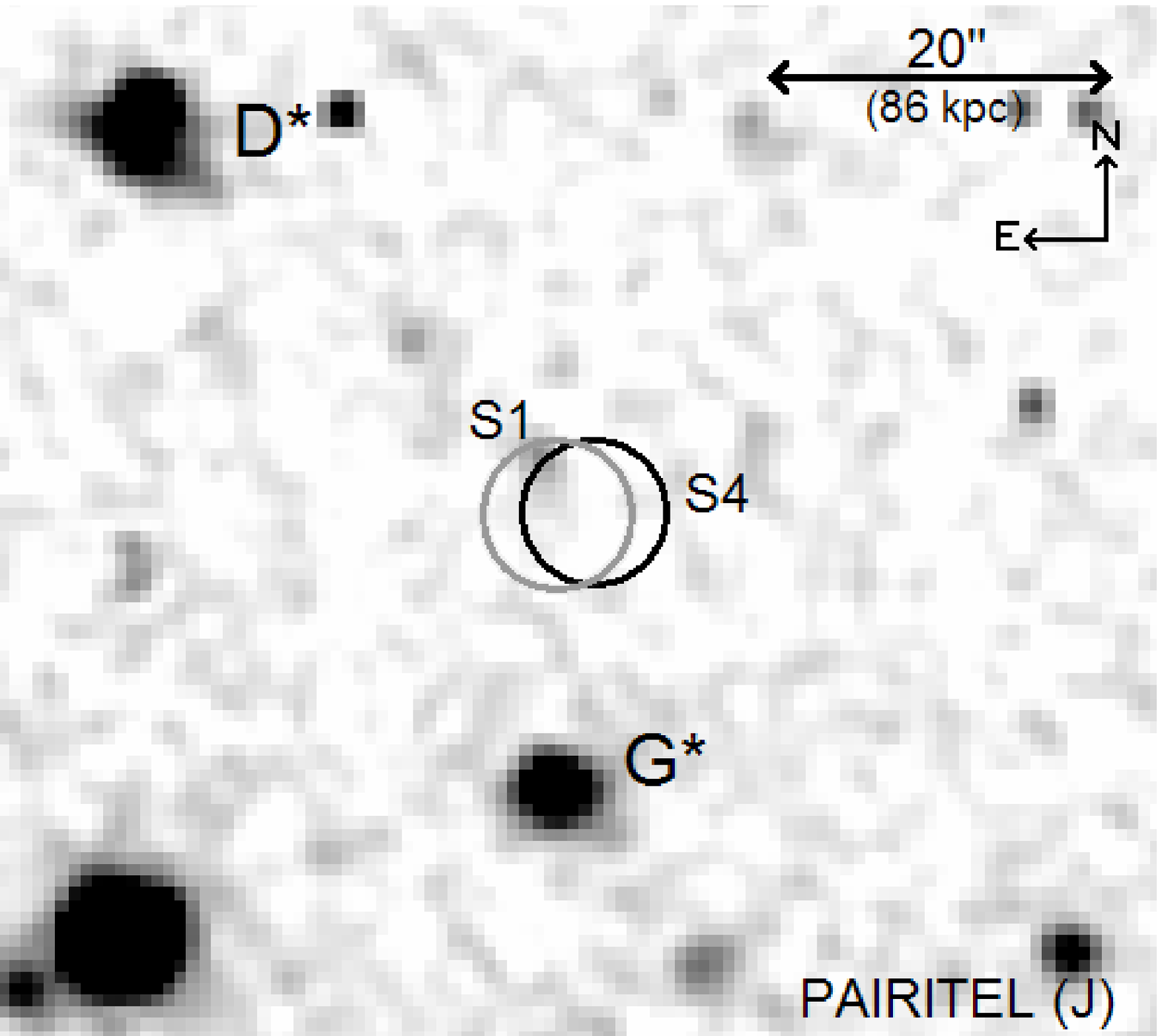,width=4.0in,angle=0}}
\caption[relation] {\small Finding chart of the field of GRB\,060502b
from Keck ({\it above}; $g'$ band) and PAIRITEL ({\it below};
$J$-band). Noted are sources discussed in the literature and in this
paper that are consistent with the Swift XRT error circle
(left; \citealt{tbg06}) and with our revised XRT error circle (right); S1
is a Galactic star. Also marked is $G^*$, which we identify as the
putative host of 060502B. The position of $G^*$ with respect the the
USNO B1.0 catalog is $\alpha$ = 18\fh35\fm45\fs.80, $\delta$ = +52\fd
37\arcmin\ 35\arcsec.9 (J2000).}
\label{fig:finder}
\end{figure*}

On 2006 May 30 {\sc UTC}, using the Low-Resolution Imaging
Spectrograph \citep{occ+95} on the Keck I 10 m telescope, we imaged
the field of SHB\, 060502B in the $R$ and $g'$ filters for 300 and 330
s, respectively. The images were processed in the usual manner. We
also observed the field from 2006 May 3 7:48:22 to 9:47:05 {\sc UTC}
with the 1.3m Peters Automated Infrared Imaging Telescope (PAIRITEL)
\citep{bsb+06}. We reduced and stacked the images in $J$, $H$, and
$K_s$ band using the standard pipeline. A Keck and PAIRITEL finding
chart of the field is presented in Figure \ref{fig:finder}.
Astrometry was performed on all images relative to the USNO B1.0
catalog \citep{mlc+03} with typical 1 $\sigma$ rms relative to that
catalog of 250 mas in each coordinate.

From the PAIRITEL imaging, we took note of an extended red source
($G^*$) to the south of the XRT position, at a position of $\alpha$ =
18\fh35\fm45\fs.76, $\delta$ = +52\fd 37\arcmin\ 36\arcsec.7
(J2000). Motivated by the inference of old galaxies at low redshift
($z \sim 0.2$) associated with some SHBs at large projected offsets
\citep{bp06} we investigated the nature of $G^*$. Photometry from the
Keck data were performed using observations of the standard star field
PG 2213 \citep{lan92}.  For $G^*$, we use a 6\arcsec\ (radius)
aperture, while a smaller aperture of 1.4\arcsec\ was used for
photometry of several fainter objects in and around the XRT error
circle.  PAIRITEL data were photometered relative to the 2MASS
\citep{scs+06} using a 6\arcsec\ radius aperture, to capture most of
the flux of $G^*$.  A summary of the photometry of this object is
found in Table \ref{tab:gphot}.

We further investigate the nature of $G^*$ by fitting different
profiles to our Keck imaging of the galaxy using the software package
GALFIT \citep{phi+02}.  Initially we fit a bulge+disk model, modeling
the galaxy as a sum of an exponential profile and a de Vaucouleurs
profile.  The de Vaucouleurs component was reduced to a point-source
by the fit, and the residuals were very large.  The residuals for a
fit with a single, general S\'{e}rsic profile were also unacceptable.
A significantly better fit was obtained with a model of the sum of two
general S\'{e}rsic profiles; the best fit for this model is an inner
component with half-light radius $R_s$ = 0.56\arcsec\ and S\'{e}rsic
index $n$ = 0.82 (approximately exponential) and a very sharp, nearly
box-car outer component with $R_s$ = 2.47\arcsec\ and $n$ = 0.12.  The
residuals have a spiral arm appearance in $g'$-band; these features
are not detected in residual fits in the R-band image, suggestive of
blue color and likely some star formation.  The degree of
concentration (low $n$ of both fits) is surprising given the red color
and small amount of star formation in this galaxy.  However, we note
that (1) this may be to some degree an overestimate of the
concentration, which has been shown \citep{bhb+03} to be
seeing-dependent, and (2) while commonly associated in older
literature only with slow-decaying profiles such as de Vaucouleurs
($n$ = 4), large surveys have shown that old, red galaxies exhibit
a wide range of profile indices from $<$ 1 up to 5 \citep{bhb+03}, and
concentrated profiles are not necessarily surprising.

\subsection{Spectroscopy}

On 2006 May 31 {\sc UTC}, we obtained spectra of $G^*$ using a 1.0\arcsec\ slit
at an angle of 15.7 East of North to also include the nearby faint
galaxy `G1' in the slit. Several spectrophotometric standard stars were
observed throughout the night at different airmasses. Spectra of $G^*$
were obtained at a median airmass of 1.37. At this angle and with this
airmass, the differential slit losses are expected to be considerable,
so we correct our resultant spectra using the broadband photometry as
described above.

\begin{figure*}[p] 
\centerline{\psfig{file=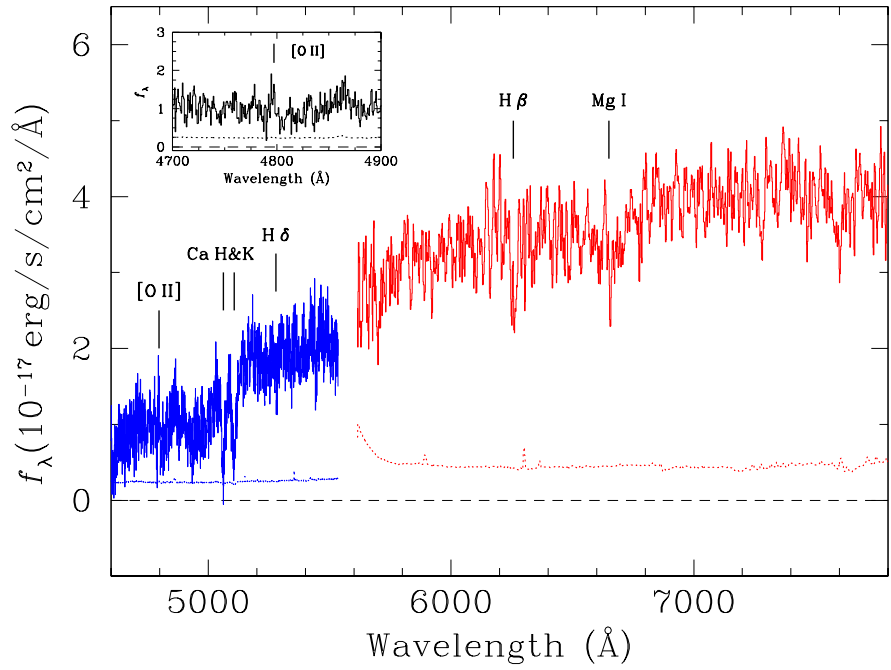,width=5.5in,angle=270}}
\caption[relation] {\small Observed Keck spectrum of $G^*$ from the blue and redside chips of LRIS. A model for the rms noise is shown below the spectrum. Prominent absorption and emission features are labeled.}
\label{fig:keckobs}
\end{figure*}

The spectrum of $G^*$ (Figure \ref{fig:keckobs}) exhibits prominent
absorption features due to the Ca\,II H+K doublet and the hydrogen
Balmer series, as well as a weak emission line due to [O\,II] at
$z=0.287$.  These spectral signatures suggest that galaxy $G^*$ is a
post-starburst system with a small amount of on-going star formation.

To determine the on-going star formation rate, we measure an
equivalent width of $W_{\rm obs}= 4.1\pm 0.7$ \AA\ for the [O\,II]
line.  Given a significant differential slit loss, we scale the
spectral continuum to match the observed broad-band flux in the $g'$
band, $AB(g')=20.12 \pm 0.03$ (corrected for Galactic extinction), and
estimate a total line flux of $f({\rm [O\,II]})=(2.3 \pm 0.4)\times
10^{-16}$ erg s$^{-1}$ cm$^{-2}$.  At $z=0.287$, the observed line
flux corresponds to a total luminosity of $L({\rm [O\,II]})=(6.1\pm
0.9)\times 10^{40}$ erg s$^{-1}$.  This indicates an on-going star
formation rate of $\sim 0.8$ M$_\odot$ yr$^{-1}$, following the
empirical relation of \citet{ken98}, or 0.4 M$_\odot$ yr$^{-1}$,
following \citet{kgj04} with no extinction correction for the observed
$L({\rm [O\,II]})$.

\begin{figure*}[p] 
\centerline{\psfig{file=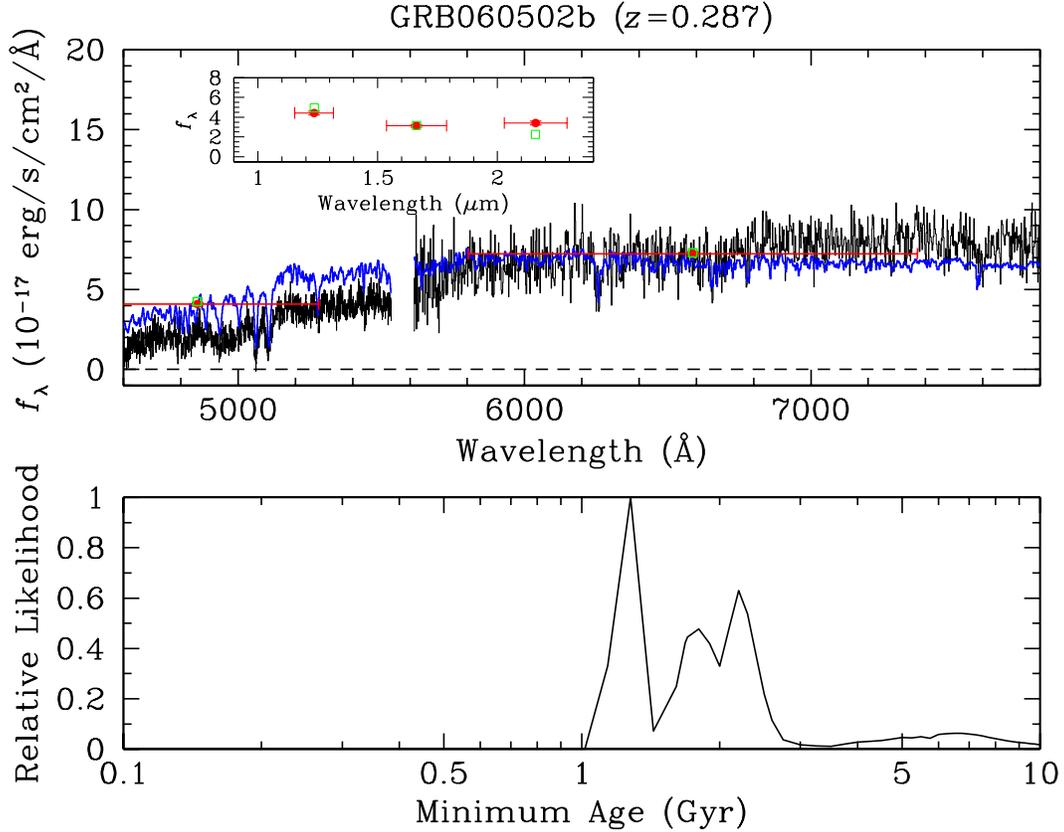,width=5.5in,angle=270}}
\caption[relation] {\small A model of the galaxy age and star formation history of $G^*$. ({\it Top}) The model (blue curve and green squares) overplotted on the Keck spectrum (black) and (red) broadband photometry corrected for Galactic extinction of $A_V = 0.146$\, mag \citep{sfd98}. Inset is the near IR photometry both observed and modeled. ({\it Bottom}) The inferred star formation history from the model, indicating a recent starburst about 1.3 Gyr prior to the GRB explosion and some extended star formation activity prior.}
\label{fig:model}
\end{figure*}

To constrain the underlying stellar population, we consider a suite of
synthetic stellar population models generated using the \citet{bc03}
spectral library.  We adopt a Salpeter initial mass function with a
range of metallicity from 1/5 solar to solar and a range of star
formation history from a single burst to an exponentially declining
star formation rate of e-folding time 300 Myr.  We include no dust in
our synthetic spectra.  Comparing the observed narrow-band features
and broad-band photometry with model predictions allows us to
constrain the stellar age in galaxy $G^*$.  The results are presented
in Figure \ref{fig:model}, where the observed spectral energy
distribution of the galaxy is shown in the top panel together with the
best-fit model.  The bottom panel of Figure \ref{fig:model} shows the
likelihood distribution function versus stellar age, indicating that
the last major episode of star formation occurred at $\approx$ 1.3 Gyr
ago.

The velocity dispersion along the slit angle was 460 km s$^{-1}$,
suggesting that $G^*$ is a massive galaxy. The absolute $K$-band
magnitude is $M_K \approx -23.3$\, mag, implying it is 1.6 $L_*$ for
early-type galaxies \citep{kpf+01}. The best-fit stellar population
and age suggest $M/L_K = 2.2$, leading to a total stellar mass of $7
\times 10^{11} h^{-2}_{71} M_{\sun}$. With a restframe equivalent
width of H$\delta$ $W_{\rm rest}= 2.9 \pm 0.5$ \AA\ and [OII] $W_{\rm
rest}= 3.2\pm 0.5$ \AA, the classification of $G^*$ is closest to `k',
though could be consistent with `k+a' (formerly part of the `E+A'
class) (following Fig.~4 of \citealt{dsp+99}).

\section{$G^*$ as the host of SHB 060502B}
\label{sec:host}

We advance the hypothesis that $G^*$ hosted the birth of the
progenitor of SHB\, 060502B, which travelled an appreciable distance
from its birthsite before producing the GRB event. At a redshift of
$z=0.287$, the offset of the XRT position from the center of the
galaxy ($17.05\arcsec \pm 4.36$\arcsec) corresponds to $r = 73 \pm 19$
$h_{71}^{-1}$ kpc in projection. With a fluence of $(4.0 \pm 0.5)
\times 10^{-8}$ erg cm$^{-2}$, the total energy release in
$\gamma$-rays assuming a unity $k$-correction is $E_{\rm iso, \gamma}
= (79 \pm 10) \times 10^{47}$ erg. Our proposed association is based
on both probabilistic and associative grounds. This is supported by a dynamical calculation in \S \ref{sec:conc}.
 
\subsection{Probabilistic Arguments}

Even with the relatively large offset observed between GRB\, 060502B
and $G^*$, the rarity of bright galaxies on the sky suggests an
association.  Based on our PAIRITEL photometry, the putative host
$G^*$ has an apparent magnitude of $K$ = 15.23.  The sky density of galaxies in the infrared at magnitudes $K<$ 18 follows the approximate distribution $d\sigma/dm = 160\times10^{0.6\,(K-15)}$ mag$^{-1}$ deg$^{-2}$ \citep{kpf+01}. Integrating this distribution, we calculate a sky density of about 250 galaxies per square degree of equal or greater brightness to $G^*$.  The probability of a given GRB occurring by chance within the observed offset of 20\arcsec\ (using the far edge of the XRT error circle) from the center of such a galaxy is $\approx$ 0.025.  Using the galaxy counts from the Calar Alto Deep
Imaging Survey \citep{htk+01} we calculate similar probabilities with
the $R$-band magnitude ($P$[$R<$ 18.5] $\approx$ 0.03) and $B$-band
magnitude ($P$[$B<$ 20.5] $\approx$ 0.05).  These estimates do not
consider host type or classification, the inclusion of which would generally serve to lower the probability. We address other galaxies as potential hosts in \S \ref{sec:other}.

\subsection{Associative Connection}

Probabilistic arguments aside, there are some heuristic arguments for
the connection worth noting. Though not all SHBs have been associated
with early-type galaxies, there is growing evidence from the small
sample that SHBs are associated with older stars  \citep{ngpf06,pbc+06,gp06,zr06}, and likely with larger burst offsets from galaxies, than LSBs \citep{bp06}. Such a
configuration is natural in the degenerate merger models; in
particular, 75 kpc offsets from massive galaxies were predicted from
{\it ab initio} binary evolution studies \citep{fwh99,bsp99}. As such,
we contend that there is now {\it a priori} precedent to support our
claim that $G^*$ hosted the progenitor birth of SHB\, 060502B.

The $G^*$ -- SHB\,060502B configuration shows some striking
similarities with the other SHBs with putative early-type hosts (see
Table \ref{tab:gang}). Not only would the redshift of $z=0.287$ be
remarkably similar to that of SHB\, 050509b and SHB\, 050724 but the
inferred energy would be consistent with that of 050509b. Indeed in
energy, redshift, putative galaxy color and type, and offset scale,
SHB\,060502b finds a strong analog in SHB\,050509b. Last, we note that
the weak X-ray afterglow and no detected optical afterglow would seem
to indicate a low density circumburst environment, as would be
expected if the GRB originated far from the progenitor birthsite.

\subsection{Consideration of Other Potential Hosts}
\label{sec:other}

With such a large physical offset, the possibility remains that the association with the putative host is coincidental and in fact the GRB originates from a different source.  Here we discuss a few alternative possibilities for the host galaxy of this GRB.  While none of these possibilities can be strongly ruled out, we nevertheless consider them less likely as potential hosts than $G*$, for various stated reasons.

The original XRT error circle contains two other optical sources, designated 'G1' and 'S2'.  Our refined XRT error circle, while generally consistent with the original XRT error circle, excludes both of these sources to 90\% confidence.  Nevertheless, as this does not completely eliminate the possibility of association (especially considering the possibility of ejection), we can ask whether or not the proximity of these sources to the XRT position suggests, on probabilistic grounds, that one of these objects is physically associated with the GRB.  The extended object G1 (the brightest source and therefore the least likely to be coincident with the error circle by random chance) has a magnitude of $R \approx$ 24; the integrated sky density for galaxies of equal or greater brightness is about 20 per arcmin$^2$.  The probability of a chance association with such an object at this distance or less is $\approx$ 0.5 -- that is, a randomly placed XRT error circle of this size will be as close or closer to such a galaxy about half the time.  The probabilities will be comparable or higher for S2 and several additional, fainter sources we identify in our imaging (S3, S4, and G2).  So while association of the GRB with one of these faint sources cannot be ruled out, the large size of the XRT error circle simply does not allow this possibility to be strongly tested.

Visible on our LRIS imaging is a nearly edge-on spiral at a distance of 34\arcsec\ northwest of the center of the XRT error circle.  Unfortunately this galaxy is strongly blended with a bright Galactic star, so an accurate magnitude measurement is difficult, though the blended source has a combined magnitude of 15.56 in the 2MASS catalog, slightly fainter than $G^*$.  Even making the conservative assumption that $K_{D*} \approx K_{G*}$, however, the probability of random association with an object of this magnitude at this distance is about a factor of 4 larger than for the association with $G^*$.   So on probabilistic grounds, if we are to associate GRB 060502B with any object in Figure 1, $G*$ is by far the strongest candidate.

There are two additional objects visible at much greater angular distances from the GRB that suggest themselves as possible hosts on account of their unusual brightness.   At a distance of 2.0\arcmin\ north of the XRT position is a bright spiral galaxy, visible in 2MASS with a magnitude of $K = 12.5$.  Despite this large distance, the probability of 'random' association in this case is $\approx 0.043$, about twice that of association with $G*$.  Even more suggestively, at a distance of 6.9\arcmin\ is the bright galaxy UGC 11292, and with a magnitude of $K$ = 10.05 \citep{kpf+01}, the probability of such a close random association is only 0.005 (less than our probability for $G*$).  Still, we tend to disfavor this hypothesis on theoretical grounds:  at the measured redshift of this galaxy of $z = 0.0276$ \citep{kpf+01}, the physical offset between the galaxy center and the XRT position is 230 kpc.  UGC\, 11292 is a very massive galaxy (at $M_K = -25.4$ it is probably several times as massive as $G*$).  Even with conservative assumptions about the galaxy mass and the position of the progenitor birthplace within it, a large kick ($v \age$ 500 km s$^{-1}$) would be required to eject an object to this distance\footnotemark\footnotetext{To be sure, systemic kicks of $>$ 500 km s$^{-1}$ are expected for compact object binaries (e.g., NS--NS, or NS--BH binaries), but most systems in population synthesis studies receive lower-velocity kicks ($v \approx 100$ km s$^{-1}$); thus, the {\it prior} expectation, in choosing between two possible kick velocities, would be weighted towards the smaller of the two inferred velocities. }.  An intriguing alternative possibility might be that the GRB was ejected from a much smaller and much less notable host that itself is associated with UGC 11292.  Perhaps the spiral galaxy mentioned above is a member of such an association; its gravitational potential well would be much more shallow and the offset would be only $\sim$70 kpc.  This is within the range of predicted short-hard GRB offsets.  However, although there is some evidence that some short-hard GRBs may originate from the local universe \citep{tcl+05}, no specific short GRB has yet been associated with any host with $z < 0.2$.  Until the local population of short GRBs and their hosts (if real) has been better characterized or other low-probability chance associations with nearby galaxies are observed, this alternative hypothesis remains extremely speculative, and the {\it a posteriori} probability argument alone is not sufficient to consider UGC\, 11292 or its hypothetical group a likely host.

\section{Discussion and Conclusions}
\label{sec:conc}

The large offset from what we have argued is a plausible host, if
true, holds important ramifications for both the sort of viable
progenitors and where they are born. First, the large offset would
seem to be at odds with the hypothesis of a degenerate binary origin
where systematic kicks are small (such as in globular clusters [GCs];
\citealt{gpm06}). While the expected number density of GCs at 75 kpc
is exceedingly small \citepeg{bbb+05}, there certainly could be a GC
at $z = 0.287$ in the XRT error circle (it would appear as faint red
point source with magnitude $R \approx 29$, in principle observable
with HST imaging). Alternatively, $G^*$ could have undergone a major
merger leaving behind a progenitor system at the XRT position. Second,
if the progenitor was created during what appears to be the last
starburst in the putative host, then the time since zero age main
sequence would be $\tau \approx 1.3 \pm 0.2$ Gyr (90\% confidence). At
the inferred offset, this would imply a minimum systemic kick velocity
of $v_{\rm kick, min} = r/\tau \approx 55 \pm 14$ km s$^{-1}$. Such a
kick velocity is comparable to the models for degenerate binaries
\citep{fwh99} and observations of Galactic double NS systems
\citep{dpp05}. The kick could have been significantly larger, implying
that the progenitor orbited about the host before the GRB
event. Indeed with the inferred stellar mass $7 \times 10^{11}
M_{\sun}\, h_{71}^{-2}$ of the putative host, unless the progenitor
was born on the outskirts of the host gravitational potential, the
true $v_{\rm kick}$ would have to have been comparable to or greater
than dispersion velocity of the host.

If the progenitor remains gravitationally bound to $G^*$ then the
systemic orbital velocity of progenitor spends most time near zero velocity, with its
initial kinetic energy stored as gravitational potential. That is, we
nominally expect an orbiting progenitor to produce a burst near the
maximal distance from its host. Indeed if all the energy is stored as
potential, then for SHB 060502B, the gravitational potential of the
progenitor system is
$$
\epsilon_{\rm pot} = \frac{G\, M_{G^*}}{d} \approx 6 \times 10^{14} ~~{\rm erg~gm}^{-1} 
	\left(\frac{M_{G^*}}{10^{12} M_\odot}\right)
	\left(\frac{d}{73~{\rm kpc}}\right)^{-1}.
$$
Upon birth, the kinetic energy per unit mass imparted to the progenitor must have been:
$$
\epsilon_{\rm kin} = \frac{1}{2} v_{\rm kick}^2 \approx 1 \times 10^{14} ~~{\rm erg~gm}^{-1} \left(\frac{v_{\rm kick}}{160\, {\rm km~s}^{-1}}\right)^2.
$$
Here we have taken the nominal velocity of the kick as the geometric mean of the dispersion velocity ($\approx 460$ km s$^{-1}$) and $v_{\rm kick, min} $; that is, we assume $v_{\rm kick} = 160$ km s$^{-1}$. That $\epsilon_{\rm kin}$ is even within an order of magnitude of $\epsilon_{\rm pot}$ is either a remarkable coincidence\footnotemark\footnotetext{If the progenitor is a double neutron star with $M_{\rm prog} = 2.8 M_\odot$, then the total respective energies are $E_{\rm pot} = M_{\rm prog} \epsilon_{\rm pot} = 3 \times 10^{48}$ erg and  $E_{\rm kin} = M_{\rm prog} \epsilon_{\rm kin} = 7 \times 10^{47}$ erg. We can think of no progenitor model to explain why $E_{\rm kin}$  and  $E_{\rm pot}$ is also comparable to $E_{\rm iso, \gamma}$.} or, we suggest, indicative of support on dynamical grounds for the ejection hypothesis.

We end by acknowledging the difficulty of confirming, beyond reasonable
doubt, our hypothesis that $G^*$ hosted the birth of the progenitor of
SHB\,060205b. The progenitors of most LSBs, owing to their connection
with massive stars, allowed for unambiguous associations with putative
hosts --- most with probability of chance alignment $P \ale 10^{-3}$
\citep{bkd02}. With SHB\,060502b we have estimated under mildly
conservative assumptions (ie.\ without regard to host type) that the
chance of a spurious assignment with $G^*$ is $P \ale$ 10\%. The
$\epsilon_{\rm kin} \approx \epsilon_{\rm pot}$ argument and the
similarity with GRB\, 050509b likely strengthen this particular
association. Yet with SHBs, especially if the majority of progenitors
are long-lived high-velocity degenerate mergers, the community must
accept that an appreciable fraction of host assignments relative to
LSBs will be spurious \citep{btw97}. Of course absorption line
redshifts of SHB afterglows, one of the remaining observational goals
of the field, will help to significantly cull the number density of
viable hosts on the sky.

\acknowledgements

We thank R.~Kennicutt and A.~Dressler for helpful discussions about
the nature of $G^*$. JSB thanks the J.\ M.\ Paredes and the Univ.\ of
Barcelona as hosts. He also acknowledges fruitful conversations with
J.\ Horth, D.\ Watson, \& J.\ Fynbo during his stay at the Dark
Cosmology Centre in Copenhagen. We thank the referee for a careful reading and thoughtful comments on the submitted manuscript. The Peters Automated Infrared Imaging
Telescope (PAIRITEL) is operated by the Smithsonian Astrophysical
Observatory (SAO) and was made possible by a grant from the Harvard
University Milton Fund, the camera loan from the University of
Virginia, and the continued support of the SAO and UC Berkeley. The
PAIRITEL project and DP are further supported by NASA/Swift Guest
Investigator Grant \# NNG06GH50G. We thank M.\ Skrutskie for his
continued support of the PAIRITEL project. Some of the data presented
herein were obtained at the W.\ M.\ Keck Observatory, which is
operated as a scientific partnership among the California Institute of
Technology, the University of California, and NASA; the Observatory
was made possible by the generous financial support of the W.\ M.\
Keck Foundation. We wish to extend special thanks to those of Hawaiian
ancestry on whose sacred mountain we are privileged to be guests.


\begin{deluxetable}{ll|ll}
\tablewidth{0pc}
\tablecaption{Photometry of the Putative Host of GRB\, 060502B\label{tab:gphot}}
\tablehead{
 \colhead{Filter} & \colhead{Magnitude}  & \colhead{Instrument/Survey} & \colhead{Reference}}
\startdata
 $B$  & 19.7          & USNO-A    & \cite{mcd+98}\\
 $B$  & 20.40         & APM-North & \cite{mim00}. \\
 $B$  & 20.58         & USNO-B    & \cite{mlc+03}\\ 
 $B$  & 19.75         & USNO-B    & \cite{mlc+03}\\
 $g'$ & 20.290 $\pm$ 0.01 & Keck-LRIS & This work \\
 $R$  & 18.6          & USNO-A    & \cite{mim00}\\
 $R$  & 18.62         & USNO-B    & \cite{mlc+03}\\
 $R$  & 18.21         & USNO-B    & \cite{mlc+03}\\
 $R$  & 18.06         & APM-North & \cite{mim00}\\
 $R$  & 18.5          & GSC2.2    & \cite{mgl+02}\\ 
 $R$  & 18.711 $\pm$ 0.01 & Keck-LRIS & This work \\
 $I$  & 17.77         & USNO-B    & \cite{mlc+03}\\
 $J$  & 17.16 $\pm$ 0.05 & PAIRITEL & This work \\
 $H$  & 16.43 $\pm$ 0.05 & PAIRITEL & This work \\
 $K_s$& 15.23 $\pm$ 0.05 & PAIRITEL & This work \\
\enddata
\end{deluxetable} 

\begin{deluxetable}{lll}
\tablewidth{0pc}
\tablecaption{Keck Photometry of faint objects in or near the XRT error circle\label{tab:xrtphot}}
\tablehead{
 \colhead{Object} & \colhead{$g'$}  & \colhead{$R$} \\
  \colhead{} & \colhead{[mag]} & \colhead{[mag]}}
\startdata
S1 & 23.438 $\pm$ 0.03 & 21.756 $\pm$ 0.01 \\
G1 & 25.937 $\pm$ 0.07 & 24.028 $\pm$ 0.06 \\
S2 & 26.557 $\pm$ 0.12 & 26.049 $\pm$ 0.27 \\
S3 & 26.561 $\pm$ 0.12 & 26.480 $\pm$ 0.52 \\
S4 & 26.799 $\pm$ 0.15 & 26.219 $\pm$ 0.41 \\
G2 & 27.944 $\pm$ 0.40 & $>$ 26.5 \\  
\enddata
\end{deluxetable}

\begin{deluxetable}{lcccr}
\tablewidth{0pc}
\tablecaption{Gang of Three: The inferred properties of SHBs associated with Early-Type Galaxies\label{tab:gang}}
\tablehead{
 \colhead{SHB} & \colhead{$z$}  & \colhead{$E_{{\rm iso}, \gamma}$} &  \colhead{$d_{\rm proj}$\tablenotemark{a}} & \colhead{Refs.} \\
\colhead{} & \colhead{} & \colhead{[erg]} & \colhead{[kpc]}}
\startdata
050509b & 0.225 & $(27 \pm 10) \times 10^{47}$ & 39 $\pm$ 13   & \citet{bpp+06} \\
050724  & 0.258 &    $1.0 \times 10^{50}$      & 2.4 $\pm$ 0.9 &  \citet{pbc+06}\\
060502b & 0.287 & $(79 \pm 15) \times 10^{47}$ & $73 \pm 19$   & This work
\enddata
\tablenotetext{a}{Projected physical offset from putative host in units of $h_{71}^{-1}$.}
\tablecomments{We do not include SHB 050813 owing to an uncertain redshift and uncertain association with a galaxy.}

\end{deluxetable} 

\end{document}